\documentclass[seceq]{ptptex}
\usepackage{wrapft}

\newcommand{\half}{ \frac{1}{2} }

\newcommand{\ket}[1]{\big|~#1~\big\rangle}

\newcommand{\contraction}[1]{\big\langle~#1~\big\rangle }

\title{
Spectrum-Generating Algebra for Charged Superstrings in Background Gauge Fields

}

\author{
Akira \textsc{Kokado}$^{1,}$\footnote{E-mail: kokado@kobe-kiu.ac.jp},  
Gaku \textsc{Konisi}$^{2,}$\footnote{E-mail: konisi@womat.zaq.ne.jp} 
and Takesi \textsc{Saito}$^{2,}$\footnote{E-mail: tsaito@k7.dion.ne.jp}
}

\inst{
$^1$Kobe International University, Kobe 658-0032, Japan\\
$^2$Department of Physics, Kwansei Gakuin University, Sanda, 669-1337, Japan
}

\abst{
The spectrum-generating algebra (SGA) for charged superstrings placed in constant background magnetic fields is constructed. Contrary to the neutral string this algebra is characteristic of including the cyclotron frequency. The Regge intercept and superconformal anomalies are calculated to be shifted by the cyclotron frequency and $D$-brane parameters from the conventional values.}

\begin{document}

\maketitle
\section{Introduction}
It is well known that in the theory of string the electromagnetic field couples with charges located only at both ends of the open string. This comes from the fact that the system should be conformally invariant \cite{ref:Matsuda}. Both charges are allowed generally to take different values with each other. Abouelsaood et al \cite{ref:Abouelsaood} first discussed the spectrum of such a charged open bosonic string in a background gauge field. They showed that in the limit of a constant field, the system is exactly solvable to give the corresponding Virasoro algebra and energy spectrum including the cyclotron frequency. \\
\indent
This model was applied to finding the pair production rate of charged strings in the presence of the constant electric and magnetic fields\cite{ref:Bachas}. Recently, there has been  again a revival of this model in the noncommutative  $D$-brane physics for strings with neutral and charged \cite{ref:Seiberg}. In a previous paper \cite{ref:Kokado1} we took the charged bosonic string, and constructed the spectrum-generating algebra (SGA) for this system. This algebra is characteristic at the point that it includes the cyclotron frequency. The SGA plays an important role when constructing the physical states satisfying the Virasoro conditions \cite{ref:Brower1}\cite{ref:Brower2} or the BRST conditions.\\
\indent
The aim of the present paper is to extend the previous work \cite{ref:Kokado1} to the charged open superstring in constant magnetic fields. We construct the super SGA for this system, and find that the Regge intercept and superconformal anomalies are shifted by the cyclotron frequency and $D$-brane parameters from the conventional values.\\
\indent
In \S2 we consider supersymmetric boundary conditions appropriate to the charged superstring in constant magnetic fields. In \S3 and \S4 we calculate the anomalies and the Regge intercepts for the present model. In \S5 the super SGA for the charged superstring is constructed. In \S6 we calculate the general formula of Regge intercept and superconformal anomalies for two different $D$-branes, where the charged open string is attached to the $Dp$-and $Dp'$-branes at $\sigma =0$ and $\sigma =\pi $, respectively. The final section is devoted to concluding remarks.

\section{Supersymmetric boundary conditions} \label{sec:2}
\indent
We consider a system in which the constant electromagnetic field couples to charges $q_0$ and $q_\pi $ at the ends of the open superstring.  The supersymmetric action for such a system is usually given by \cite{ref:Bachas}, \cite{ref:Seiberg} 
\begin{eqnarray}
  S=-\frac{1}{2} \int d^2 \sigma \big( \partial _\alpha X^\mu \partial^\alpha X_\mu - i \bar{\psi }^\mu \gamma ^\alpha \partial_\alpha \psi_\mu  \big) \nonumber \\ 
  - \frac{1}{2}\sum_{\sigma  =0,\pi } \int d \tau q_\sigma \big( \dot{X}^\mu F_\mu^{\ \nu } X_\nu 
  + \frac{i}{2} \psi^\mu F_\mu^{\ \nu } \psi_\nu \big)~,
\label{eq:action1} 
\end{eqnarray}
where
\[
  \gamma ^0
  = \begin{pmatrix}
              0 &  -i \\
              i & 0
    \end{pmatrix}~, \qquad 
  \gamma ^1
  = \begin{pmatrix}
              0 &  i \\
              i & 0 
    \end{pmatrix}~,  \qquad
  \{\gamma ^\alpha , \gamma ^\beta \} = -2\eta ^{\alpha \beta }~, \qquad
  \eta ^{00}=-1~,  
\]
\[
  \bar {\psi } = \psi ^{T\mu }\gamma ^0~, \qquad 
  \psi ^\mu 
  = \begin{pmatrix}
              \psi _1^{\ \mu } \\ 
              \psi _2^{\ \mu } 
    \end{pmatrix}~. 
\]
Here we have already taken a gauge $A_\mu (X)=-(1/2)F_\mu ^{\ \nu }X_\nu $ with $F_\mu ^{\ \nu }$ constant. The last term shows that the fermionic field interacts only with the field strength in the Pauli type, where the intrinsic spin operator is given by $S^{\mu \nu }=[\psi ^\mu ~, \psi ^\nu ]/2i$.\footnote{
We can introduce the NS antisymmetric background field to the action (\ref{eq:action1}) together with $F_\mu^{\ \nu }$.  However, the action is not modified if we use the rescaled fields and  the rescaled charges.}

   From this action we have equations of motion
\begin{eqnarray}
   \partial _\alpha  \partial ^\alpha  X^\mu (\tau ~, \sigma~) = 0~, 
   \label{eq:motion_of_X} \\
   \gamma _\alpha \partial ^\alpha \psi ^\mu (\tau ~, \sigma~) = 0~.
    \label{eq:motion_of_psi} 
\end{eqnarray}
with boundary conditions at $\sigma =0,\pi $
\begin{eqnarray}
   \partial _\sigma  X^\mu (\tau ~, \sigma ~) = -\rho (\sigma )F_\mu ^{\ \nu }\partial_\tau  X^\nu (\tau ~,\sigma ~)~, \quad
   \textrm{or in short} \quad X'=-\rho (\sigma )F\dot{X}~, 
  \label{eq:bc_X} \\
   \Delta \psi _1^{\ \mu }\big[ {\psi _1}_\mu - \rho (\sigma ) F_\mu ^{\ \nu } {\psi _1}_\nu \big]
   - \Delta \psi _2^{\ \mu }\big[{\psi _2}_\mu  + \rho (\sigma ) F_\mu ^{\ \nu } {\psi _2}_\nu \big]=0~.
  \label{eq:bc_psi}
\end{eqnarray}
where $\rho (0)=q_0$, $\rho (\pi )=-q_\pi $, and $\Delta \psi '$s are small variations of $\psi '$s. These boundary conditions provide primary constraints and necessitate the Dirac quantization. We first do not consider any $D$-branes until Sec.6.\\
\indent
The fermionic boundary conditions are usually taken as
\begin{eqnarray}
   \big[{\psi _1}_\mu - \rho (\sigma ) F_\mu ^{\ \nu }{\psi _1}_\nu \big] -
   \big[{\psi _2}_\mu + \rho (\sigma ) F_\mu ^{\ \nu }{\psi _2}_\nu \big] =0~, \qquad (\sigma =0)
   \label{eq:bc_of_NS} \\
   \big[{\psi _1}_\mu - \rho (\sigma ) F_\mu ^{\ \nu }{\psi _1}_\nu \big] +
   \big[{\psi _2}_\mu + \rho (\sigma ) F_\mu ^{\ \nu }{\psi _2}_\nu \big] =0~, \qquad (\sigma =\pi )
   \label{eq:2_bd_of_psi12}
\end{eqnarray}
for the Neveu-Schwarz (NS) sector, and 
\begin{equation}
   \big[{\psi _1}_\mu - \rho (\sigma ) F_\mu ^{\ \nu }{\psi _1}_\nu \big] -
   \big[{\psi _2}_\mu + \rho (\sigma ) F_\mu ^{\ \nu }{\psi _2}_\nu \big] =0~, \qquad (\sigma =0, \pi )
   \label{eq:bc_of_R}
\end{equation}
for the Ramond (R) sector. This means that $\Delta \psi _1=\pm \Delta \psi _2=$ arbitrary, but $\psi _1 \neq \psi _2$.
 This is because Eqs.(\ref{eq:bc_X})
 and (\ref{eq:bc_of_NS})-(\ref{eq:bc_of_R}) are an invariant pair under the supersymmetric transformation. In the notation as $\psi =\psi _1(\tau -\sigma )$, $\bar {\psi }=\psi _2(\tau +\sigma )$ and $\partial X=\dot{X}-X'$, $\bar{\partial }X=\dot{X}+X'$, both boundary conditions are expressed as
\begin{eqnarray}
  (1-\rho F)\partial X = (1+\rho F)\bar{\partial }X~,
  \label{eq:2_bc_dX} \\
  (1-\rho F)\psi  = \pm (1+\rho F)\bar{\psi }~,
  \label{eq:2_bc_psi2} 
\end{eqnarray}
Both equations are exchanged with each other by the supersymmetric transformation,
\begin{eqnarray}
  \delta X &=& -i\eta (\bar{\psi }+\psi )~,
  \label{eq:supersymmetric_tr} \\
  \delta \psi &=& \eta \partial  X~,
  \nonumber \\
  \delta \bar{\psi} &=& \eta \bar{\partial }X~.
  \nonumber
\end{eqnarray}
The action is invariant under such a supersymmetric transformation. This can be proved from the conservation of the supercurrent $J^\alpha   =\gamma^\beta \gamma^\alpha \partial_\beta X_\mu \psi ^\mu $ with use of equations of motion. 
%
\section{Calculation of anomaly}  \label{sec:3}

We concentrate on one $2\times 2$ block of $F_\mu ^{\ \nu }$  
\begin{equation}
  F_\mu ^{\ \nu } = \begin{pmatrix} 0 & B \\
                             -B & 0 
                    \end{pmatrix}~, \qquad \mu ,\nu =1,2~, \label{eq:3_def_F}
\
\end{equation}
Introducing
\begin{equation}
  X^{(\pm)} = 1/\sqrt{2}(X^1\pm i X^2) = (X^{(\mp)})^*~,
  \label{eq:3_definition_Xpm}
\end{equation}
the boundary conditions (\ref{eq:bc_X}) are diagonalized
\begin{equation}
 {X'}^{(+)} = i\rho (\sigma )B \dot{X}^{(+)} \qquad \textrm{at} \ \sigma = 0, \pi~,  
 \label{eq:3_init_X_0_pi} \\
\end{equation}
The solutions of field equation (\ref{eq:motion_of_X}) satisfying the boundary conditions are
\begin{equation}
  X^{(\pm)}(\tau, \sigma)
  = x^{(\pm)} \pm i \sum_n\, \frac{ 1}{n \pm \omega}~e^{ -i (n \pm \omega) \tau}~
     \cos\big[~(n \pm \omega) \sigma \mp \pi\omega_0~\big]~ 
     \alpha^{(\pm)}_n~,
\label{eq:3_X-mode_decomposition}
\end{equation}
where
\begin{eqnarray}
  \tan \pi\omega_0 = \rho (0)B~, \qquad \tan \pi\omega_\pi = \rho (\pi) B~, \qquad \omega =\omega _0 - \omega _\pi~. 
  \label{eq:2_del_omega0_omegapi}
\end{eqnarray}
We call $\omega $ the cyclotron frequency. By definition, it follows that $-1/2< (\omega_0, \omega_\pi) < 1/2$ and hence $|\omega | < 1$. 
In the following we choose the axes 1 and 2 to have $0< \omega < 1$. \\

The Dirac quantization of this constrained system\cite{ref:Kokado2} gives the commutation relations
\begin{eqnarray}
  \big[\, \alpha^{(+)}_m\,,~\alpha^{(-)}_n~\big] &=& (m+\omega )~\delta _{m+n,0} 
  \label{eq:3_commutation_relation_alpha} \\
  \big[\, x^{(+)}\,,~x^{(-)}~\big]  &=& - \pi\frac{\cos \pi\omega_0 \sin \pi\omega_\pi}{\sin \pi \omega }
  \label{eq:3_commutation_relation_x}
\end{eqnarray}

The supersymmetric pair of the boundary conditions (\ref{eq:2_bc_dX}) and (\ref{eq:2_bc_psi2}) suggests that the fermionic field $\psi ^\mu $ should obey the same boundary conditions as bosonic ones. This leads us, in the complex variable $z=\exp (\tau + i \sigma )$,
\begin{eqnarray}
  J^{(\pm)}(z) &=& i\partial X^{(\pm)}(z) = \sum_n \alpha _n^{(\pm)} z^{-n\mp \omega -1} = z^{\mp\omega }\tilde {J}^{(\pm)}(z)~,
  \label{eq:3_mode_expansion_J} \\
  \psi ^{(\pm)}(z) &=& \sum _r b_r ^{(\pm)} z^{-r\mp \omega -1/2} = z^{\mp\omega }\tilde {\psi }^{(\pm)}(z)~,
  \label{eq:3_modo_expansion_psi}
\end{eqnarray}
where $r$ runs over half-integers in the NS sector, and
\begin{equation}
  \big\{\, b_r^{(+)}\,,~b_s^{(-)}~\big\} = \delta _{r+s,0}
  \label{eq:3_commutation_relation_b}~.
\end{equation}
In the following we concentrate ourselves in the NS sector. \\
\indent
The operator product expansions for them are given by

\begin{eqnarray}
  \tilde {J}^{(+)}(z) \tilde {J}^{(-)}(z') &=& \frac{1}{(z-z')^2} + \frac{\omega }{z'(z-z')}~,
  \label{eq:3_operator_product_J*J} \\
  \tilde {J}^{(-)}(z) \tilde {J}^{(+)}(z') &=& \frac{1}{(z-z')^2} - \frac{\omega }{z(z-z')}~,
  \label{eq:3_operator_product_J*J2} \\
  \tilde {\psi }^{(\pm)}(z) \tilde {\psi }^{(\mp)}(z') &=& \frac{1}{z-z'} ~,
  \label{eq:3_operator_product_psi*psi}
\end{eqnarray}
where the tilde operators are those with the factor $z^{\pm \omega }$ dropped from the original ones. \\
\indent
Define the super current operator by
\begin{eqnarray}	
  T_F(z) &=& i\psi ^\mu (z) \partial X_\mu (z) = \psi^\mu(z) J_\mu (z) \nonumber \\
         &=& -\psi^0 J^0 + \psi^1 J^1 + \psi^2 J^2 + \psi^3 J^3 + \cdots + \psi^{d-1} J^{d-1}~,
  \label{eq:3_def_supercurrent_operator}
\end{eqnarray}
where
\begin{equation}
  \psi^1 J^1 + \psi^2 J^2 = \psi^{(+)} J^{(-)} + \psi^{(-)} J^{(+)} =\tilde {\psi}^{(+)} \tilde{J}^{(-)} + \tilde{\psi}^{(-)} \tilde{J}^{(+)}~.
  \label{eq:3_psi_1_and_2}
\end{equation}
The other components are irrelevant for our aim, because they are free and well known. The operator product $T_F(z)T_F(z')$ yields the conformal operator $T_B(z)$, i.e.,
\begin{eqnarray}
  T_F(z)T_F(z')
     &=& \big [\tilde{J}^{(+)}(z) \tilde{\psi}^{(-)}(z) + \tilde{J}^{(-)}(z) \tilde{\psi}^{(+)}(z) \big ] 
                  \big [\tilde{J}^{(+)}(z') \tilde{\psi}^{(-)}(z') + \tilde{J}^{(-)}(z') \tilde{\psi}^{(+)}(z') \big ] 
     \nonumber \\ 
     &=& \big\langle~\tilde{J}^{(+)}(z) \tilde{J}^{(-)}(z') ~\big\rangle \big\langle~\tilde{\psi}^{(-)}(z) \tilde{\psi}^{(+)}(z') ~\big\rangle
      +  \big\langle~\tilde{J}^{(-)}(z) \tilde{J}^{(+)}(z') ~\big\rangle \big\langle~\tilde{\psi}^{(+)}(z) \tilde{\psi}^{(-)}(z') ~\big\rangle 
       \nonumber \\
      &+& \big\langle~\tilde{J}^{(+)}(z) \tilde{J}^{(-)}(z') ~\big\rangle :\tilde{\psi}^{(-)}(z) \tilde{\psi}^{(+)}(z'):
      + :\tilde{J}^{(-)}(z) \tilde{J}^{(+)}(z'): \big\langle~\tilde{\psi}^{(+)}(z) \tilde{\psi}^{(-)}(z') ~\big\rangle
      \nonumber \\
      &+& :\tilde{J}^{(+)}(z) \tilde{J}^{(-)}(z'): \big\langle~\tilde{\psi}^{(-)}(z) \tilde{\psi}^{(+)}(z') ~\big\rangle
      +  \big\langle~\tilde{J}^{(-)}(z) \tilde{J}^{(+)}(z') ~\big\rangle :\tilde{\psi}^{(+)}(z) \tilde{\psi}^{(-)}(z'): 
       \nonumber \\
     &=& \Big[\frac{1}{(z-z')^2} + \frac{\omega }{z'(z-z')}\Big]\frac{1}{z-z'} 
      + \Big[\frac{1}{(z-z')^2} - \frac{\omega }{z(z-z')}\Big]\frac{1}{z-z'} 
      \nonumber \\
      &+& \Big[\frac{1}{(z-z')^2} + \frac{\omega }{z'(z-z')}\Big] :\tilde{\psi}^{(-)}(z) \tilde{\psi}^{(+)}(z'):
      \nonumber \\
      &+& \Big[\frac{1}{(z-z')^2} - \frac{\omega }{z(z-z')}\Big] :\tilde{\psi}^{(+)}(z) \tilde{\psi}^{(-)}(z'):
      \nonumber \\
      &+& :\tilde{J}^{(+)}(z) \tilde{J}^{(-)}(z'): \frac{1}{z-z'} 
      + :\tilde{J}^{(-)}(z) \tilde{J}^{(+)}(z'): \frac{1}{z-z'}
     \nonumber \\ 
     &=& \frac{2}{(z-z')^3} +\frac{\omega }{zz'(z-z')} +\frac{2}{z-z'} T_B(z')~,
     \label{eq:3_operator_product_TT'}
\end{eqnarray}     
where
\begin{eqnarray}
   T_B &=& :\tilde{J}^{(+)} \tilde {J}^{(-)}: + \frac{1}{2}:\big(-2\frac{\omega }{z}\tilde{\psi}^{(+)} \tilde {\psi}^{(-)}
         + \partial \tilde{\psi}^{(+)} \tilde {\psi}^{(-)} + \partial \tilde{\psi}^{(-)} \tilde {\psi}^{(+)} \big): \\
       &=& :J^{(+)} J^{(-)}: + \frac{1}{2} :\big(\partial \psi^{(+)} \psi^{(-)}\big):
   \nonumber  
   \label{eq:3_def_TB}
\end{eqnarray}
Including the other components we get the conventional form
\begin{equation}
  T_B = \frac{1}{2} :J_\mu J^\mu : + \frac{1}{2} :\partial \psi_\mu \psi^\mu :~,
  \label{eq:3_conventional_form}
\end{equation}
and
\begin{equation}
  T_F(z) T_F(z') = \frac{d}{(z-z')^3} + \frac{\omega }{zz'(z-z')} + \frac{2}{z-z'}T_B(z')~.
  \label{eq:3_result_TFTF}
\end{equation}

The singular parts of operator product $T_B(z)T_B(z')$ are also calculated as follows:
For the fermionic part we have
\begin{eqnarray}
   \textrm{Fermionic part of }T_B(z) T_B(z') &=& \frac{1}{2} :\partial \psi(z) \psi(z): \frac{1}{2} :\partial \psi(z') \psi(z'): 
   \label{eq:3_fermionic_part_TBTB} \\
   &=& :\Big[ -\omega z^{-1}\tilde {\psi}^{(+)}\tilde {\psi}^{(-)} + \frac{1}{2}\partial \tilde {\psi}^{(+)}\tilde {\psi}^{(-)}
   + \frac{1}{2}\partial \tilde {\psi}^{(-)}\tilde {\psi}^{(+)}(z) \Big]: 
  \nonumber \\
   &&:\Big[ -\omega z'^{-1}\tilde {\psi}^{(+)}\tilde {\psi}^{(-)} + \frac{1}{2}\partial' \tilde {\psi}^{(+)}\tilde {\psi}^{(-)}
   + \frac{1}{2}\partial' \tilde {\psi}^{(-)}\tilde {\psi}^{(+)}(z') \Big]:
  \nonumber \\
   &=&\frac{\omega ^2}{zz'(z-z')^2} +\frac{1}{2}\partial '\frac{1}{z-z'}\partial \frac{1}{z-z'}
   - \frac{1}{2}\frac{1}{z-z'}\partial \partial '\frac{1}{z-z'}
  \nonumber \\
   &=&\frac{\omega ^2}{zz'(z-z')^2} + \frac{1}{2(z-z')^4}
  \nonumber 
\end{eqnarray}
where the term linear to $\omega $ vanishes. On the other hand, for the bosonic part we have
\begin{eqnarray}
  \textrm{Bosonic part of } T_B(z)T_B(z') &=& :\tilde {J}^{(+)}(z)\tilde {J}^{(-)}(z)::\tilde {J}^{(+)}(z')\tilde {J}^{(-)}(z'):
  \label{eq:3_bosonic_part_TBTB} \\
  &=&\Big[\frac{1}{(z-z')^2} + \frac{\omega }{z'(z-z')}\Big]\Big[\frac{1}{(z-z')^2} - \frac{\omega }{z(z-z')}\Big]
  \nonumber \\
  &=& \frac{1}{(z-z')^4} + \frac{\omega -\omega ^2}{zz'(z-z')^2}~.
  \nonumber
\end{eqnarray}
The total sum of the singularities is given by Eqs.(\ref{eq:3_fermionic_part_TBTB})+(\ref{eq:3_bosonic_part_TBTB}), i.e.,
\begin{equation}
  T_B(z)T_B(z') = \frac{3/2}{(z-z')^4} + \frac{\omega }{zz'(z-z')^2}~.
  \label{eq:3_total_sum_singularities}
\end{equation}
It is a remarkable fact that the $\omega ^2$ singularity in the bosonic part is canceled by the $\omega ^2$ singularity in the fermionic part. The full operator product gives
\begin{equation}
   T_B(z) T_B(z') = \frac{3d/4}{(z-z')^4} + \frac{\omega }{zz'(z-z')^2} + \frac{2}{(z-z')^2}T_B(z') + \frac{1}{z-z'}\partial' T_B(z')~.
   \label{eq:3_full_op_TBTB}
\end{equation}
The superconformal algebra is completed by the operator product of $T_B(z)T_F(z')$:
\begin{equation}
   T_B(z)T_F(z') = \frac{3/2}{(z-z')^2}T_F(z') + \frac{1}{z-z'}\partial' T_F(z')~,
   \label{eq:3_superconformal_algebra}
\end{equation}
where the $\omega $ terms are completely disappeared. \\
\indent
To sum up, the operator products (\ref{eq:3_full_op_TBTB}), (\ref{eq:3_superconformal_algebra}) and (\ref{eq:3_result_TFTF}) for $T_F$, $T_B$ are closed to make the supersymmetric algebra. In the standard form they are equivalent to,
\begin{eqnarray}
  \big[\, L_m\,,~L_n~\big] &=& (m-n)L_{m+n} +A(m) \delta _{m+n,0}~,
  \nonumber \\
  \big[\, L_m\,,~G_r~\big] &=& \big(\frac{m}{2}-r\big)G_{m+r}~,
  \label{eq:3_commutation_relation_L_G} \\
  \big\{\, G_r\,,~G_s~\big\} &=& 2L_{r+s}+B(r)\delta _{r+s,0}~,
  \nonumber
\end{eqnarray}
where
\begin{eqnarray}
  A(m) &=& \frac{d}{8} (m^3-m) + m\omega~,
  \label{eq:3_del_A_B} \\
  B(r) &=& \frac{d}{2}\Big(r^2 - \frac{1}{4} \Big) + \omega ~,
  \label{eq:3_del_Br2} \\
  \omega &=& \sum_i \omega_i~.
  \nonumber
\end{eqnarray}
Here $\omega_i $ is the cyclotron frequency of the $i$-th block. The result for $A(m)$ should be compared with the bosonic case, where $A^{\rm{bosomic}}(m)=d(m^3-m)/12+m(\omega -\omega ^2)$. \cite{ref:Abouelsaood}\cite{ref:Kokado2} \\
\indent
As for the Ramond sector, it is remarkable that no cyclotron frequencies are involved in anomaly factors $A$, $B$. 
 Note that $F_0$ contains factors, $\alpha _0^{(+)}d_0^{(-)}+\alpha _0^{(-)}d_0^{(+)}$. Since $\alpha_0^{(-)}$ is the creation operator,
 the second term contradicts with the Virasoro condition $F_0 \ket{\rm{ground  state}}=0$, if $d_0^{(\pm)}$ is regarded as the Dirac $\gamma $
 matrix in the conventional way.  In the sector of the presence of magnetic fields, therefore, $d_0^{(+)}$ should be regarded as the annihilation
 operator, whereas other components $d_0^\mu $ without magnetic fields behave as $\gamma $ matrices.
  
\section{Isomorphisms}  \label{sec:4}

The spectrum-generating algebra for Neveu-Schwarz string contains the sub-algebra for the longitudinal operators constructed from oscillators with light-cone components, $X_{\pm}=(X^0 \pm X^{d-1})/\sqrt{2}$, where $d$ is the number of space-time dimensions. They are
\begin{eqnarray}
  \big[\, A_m^{+}\,,~A_n^{+}~\big] &=& (m-n)A_{m+n}^+ + m^3 \delta _{m+n,0}~,
  \nonumber \\
  \big[\, A_m^+\,,~B_r^+~\big] &=& (\frac{m}{2}-r)B_{m+r}^+~,
  \label{eq:4_commutation_relation_AB} \\
  \big\{\, B_r^+\,,~B_s^+~\big\} &=& 2A_{r+s}^+ + 4r^2\delta _{r+s,0}~,
  \nonumber
\end{eqnarray}
The detail will be summarized in the next section, but we assume that $d$ is even and there exists no magnetic field in the last ($d-1$) dimension. The algebra (\ref{eq:4_commutation_relation_AB}) is similar to the super Virasoro algebra for transverse operators 
\begin{eqnarray}
  \big[\, L_m^T\,,~L_n^T~\big] &=& (m-n)L_{m+n}^T +A^T(m) \delta _{m+n,0}~,
  \nonumber \\
  \big[\, L_m^T\,,~G_r^T~\big] &=& (\frac{m}{2}-r)G_{m+r}^T~,
  \label{eq:4_commutation_relation_L_G} \\
  \big\{\, G_r^T\,,~G_s^T~\big\} &=& 2L_{r+s}^T + B^T(r)\delta _{r+s,0}~,
  \nonumber
\end{eqnarray}
where
\begin{eqnarray*}
  A^T(m) &=& \frac{d-2}{8} (m^3-m) + 2ma + m\omega~, \\
  B^T(r) &=& \frac{d-2}{2}\Big(r^2 - \frac{1}{4} \Big) + 2a + \omega 
\end{eqnarray*}
Here the superscript $T$ means that the operators are constructed from oscillators with spacial components $\mu = 1,2,\cdots, d-2$. Explicitly,
\begin{eqnarray}
  L_m^T &=& -a \delta _{m,0} + \frac{1}{2} \sum_l \sum_{j=1}^{d-2}:\alpha _{-l}^j \alpha_{m+l}^{j}:
        + \frac{1}{2} \sum_{r,i} \Big(\frac{m}{2}+r+\omega_i \Big):b_{-r}^{i(-)} b_{m+r}^{i(+)}: 
        \nonumber \\
        &&+ \frac{1}{2} \sum_{r,i} \Big(\frac{m}{2}+r-\omega_i \Big):b_{-r}^{i(+)} b_{m+r}^{i(-)}:
        + \frac{1}{2} \sum_{r,k} \Big(\frac{m}{2}+r \Big):b_{-r}^k b_{m+r}^{k}:~,
        \label{eq:4_construction_Lm_Gr} \\
  G_r^T &=&  \sum_n \sum_{j=1}^{d-2} \alpha _{-n}^j b_{r+n}^{j}~,
  \nonumber
\end{eqnarray}
where $a$ corresponds to the Regge intercept $\alpha (0)$, the sum over $i$ is for all blocks, and the sum over $k$ is for dimensions of no magnetic fields.  The isomorphisms
\begin{equation}
  A_m^+ \sim L_m^T~, \qquad B_r^+ \sim G_r^T
  \label{eq:4_similar_A_B}
\end{equation}
are completed, if there hold
\begin{eqnarray}
  A^T(m) &=& \frac{d-2}{8} (m^3-m) + 2ma + m\omega = m^3~,
  \label{eq:4_condition_AT_BT} \\
  B^T(r) &=& \frac{d-2}{2}\Big(r^2 - \frac{1}{4} \Big) + 2a + \omega = 4r^2~. 
  \nonumber
\end{eqnarray}
These two equations are consistent to give the solution,
\begin{eqnarray}
  d &=&10~,
  \label{eq2:4_solution_d_a} \\
  a &=& \alpha (0) = \frac{1 - \omega}{2}~.
  \nonumber
\end{eqnarray}
We have seen that in the NS sector the Regge intercept $\alpha(0)$becomes less than 1/2 by the half of cyclotron frequencies. As for the R sector we note $a=\alpha _R(0)=0$.
  
\section{Spectrum-Generating Algebra}  \label{sec:5}
\indent
The isomorphisms (4.4), $A_m^+ \sim L_m^T$, $B_r^+ \sim G_r^T$ are also extended to other components interacting with magnetic field. The relevant algebra is the following: 
\begin{eqnarray}
  \big[\, \alpha _m^{(+)}\,,~\alpha _n^{(-)}~\big] &=& (m+\omega ) \delta _{m+n,0}~, \qquad 
  \big\{\, b_r^{(+)}\,,~b_s^{(-)}~\big\} = \delta _{r+s,0}~,
  \label{eq:6_relevanat_algebra} \\
  \big[\, \alpha _m^{(\pm)}\,,~L_n^T~\big] &=& (m\pm\omega ) \alpha _{m+n}^{(\pm)}~, \qquad
  \big[\, b_r^{(\pm)}\,,~L_n^T~\big] = (\frac{n}{2} + r \pm \omega ) b_{r+n}^{(\pm)}~, 
  \nonumber \\
  \big[\, \alpha _m^{(\pm)}\,,~G_r^T~\big] &=& (m\pm\omega )b_{m+r}^{(\pm)}~, \qquad
  \big\{\, b_r^{(\pm)}\,,~G_s^{T}~\big\} = \alpha _{r+s}^{(\pm)}~, \qquad
  \big[\, \alpha _m^i\,,~b_r^j~\big] = 0~, 
  \nonumber
\end{eqnarray}
We would like to seek the spectrum-generating algebra (SGA) such that
\begin{eqnarray}
  \big[\, A_m^{(+)}\,,~A_n^{(-)}~\big] &=& (m+\omega ) \delta _{m+n,0}~, \qquad 
  \big\{\, B_r^{(+)}\,,~B_s^{(-)}~\big\} = \delta _{r+s,0}~,
  \label{eq:6_relevanat_algebra2} \\
  \big[\, A_m^{(\pm)}\,,~A_n^{+}~\big] &=& (m\pm\omega ) A_{m+n}^{(\pm)}~, \qquad
  \big[\, B_r^{(\pm)}\,,~A_n^{(+)}~\big] = (\frac{n}{2} + r \pm \omega ) B_{r+n}^{(\pm)}~, 
  \nonumber \\
  \big[\, A_m^{(\pm)}\,,~B_r^{+}~\big] &=& (m\pm\omega )B_{m+r}^{(\pm)}~, \qquad
  \big\{\, B_r^{(\pm)}\,,~B_s^{+}~\big\} = A_{r+s}^{(\pm)}~, \qquad
  \big[\, A_m^i\,,~B_r^j~\big] = 0~, 
  \nonumber
\end{eqnarray}
A complete set of SGA is Eqs.(\ref{eq:6_relevanat_algebra2}) and Eqs.(\ref{eq:4_commutation_relation_AB}). The isomorphisms are the following:
\begin{eqnarray}
  A_m^{(\pm)} &\sim& \alpha _m^{(\pm)}~, \qquad B_r^{(\pm)} \sim b_r^{(\pm)}
  \label{eq:6_isomorphism} \\
  A_m^{+} &\sim& L_m^{T}~, \qquad B_r^{+} \sim G_r^T
  \nonumber
\end{eqnarray}
Any operator of SGA should be commutable with the super Virasoro operator $G_r$. 

We now construct such a SGA in the following:\\
\noindent
\textbf{SGA 1}. $\big[\, A_m^{(+)}\,,~A_n^{(-)}~\big]$ = $(m+\omega ) \delta _{m+n,0}$

Define relevant operators
\begin{equation}
   A_m^{(\pm)}=\frac{1}{2\pi i} \oint dz A_m^{(\pm)}(z)~,
   \label{eq:6_def_Am(pm)}
\end{equation}
where $A_m^{(\pm)} = [J^{(\pm)} - (m \pm \omega )\psi^{(\pm)}\psi_-]V^{m \pm \omega}$, and
\begin{eqnarray}
  V &=& :\exp [iX_-\-(z)]:~, \qquad V^n = :\exp[inX_-(z)]:~,
  \label{eq:6_def_V}  \\
  X_-(z) &=& x_- -ip_- \ln z + i \sum_{n \neq 0}\frac{\alpha _n^-}{n}z^{-n}~, \qquad p_- = 1~.
  \label{eq:6_mode_expansion_X-}
\end{eqnarray}
Here $\psi_-$, $X_-$ are light-cone variables defined by $X_{\pm}$ = $\kappa ^{\pm 1}(X^0 \pm X^{d-1})/\sqrt{2}$, with a real parameter $\kappa $. The superscripts $(\pm)$ of $X^{(\pm)} = (X^1\pm i X^2)/\sqrt{2}$  should be distinguished from the light-cone subscripts $\pm$. Note that the vertex operator $V^{m\pm \omega } = :\exp [i(m\pm \omega )X_-(z)]:$ behaves like $\sim z^{m\pm\omega}$ at $z=0$. \\
\indent
The new definition for $A_m^{(\pm)}$ and $B_r^{(\pm)}$ below should be compared with the original ones by Brower and Friedmann
\begin{equation}
  A_m^{i,\rm{BF}} (z) = \big[ J^i - m \psi^i \psi_- \big] V^m~,
  \label{eq:5_def_AmBF1}
\end{equation}
and
\begin{equation}
  B_r^{i,\rm{BF}} (z) = \big[ \psi^i \big(1+ \half \psi_- \partial \psi_- J_-^{-2} \big)J_-^{1/2} -\psi_- J^i J_-^{-1/2} \big] V^r~,
  \label{eq:5_def_AmBF2}
\end{equation}
in Ref.\cite{ref:Brower2}. \\
\indent 
It is the standard technique that the commutator can be expressed as the radial ordering path integral for the operator product
\begin{equation}
  \big[\, A_m^{(+)}\,,~A_n^{(-)}(z')~\big] = \frac{1}{2\pi i} \oint dz A_m^{(+)}(z)A_n^{(-)}(z')
  \label{eq:6_calculation_Am+,An-}
\end{equation}

Now, the operator product becomes
\begin{eqnarray}
  A_m^{(+)}(z) A_n^{(-)} &=& \big[J^{(+)}(z) - (m+\omega)\psi^{(+)}(z)\psi_-(z)\big]V^{m+\omega}(z)
   \big[J^{(-)}(z')  - (n-\omega)\psi^{(-)}(z')\psi_-(z')\big]V^{n-\omega}(z')
  \nonumber \\
  &=& \contraction{J^{(+)}(z) J^{(-)}(z')} V^{m+\omega}(z)V^{n-\omega}(z') 
  \nonumber \\
  && +(m+\omega)(n-\omega)\contraction{\psi^{(+)}(z) \psi^{(-)}(z')}\psi_-(z)\psi_-(z')V^{m+\omega}(z)V^{n-\omega}(z')
  \nonumber \\
  &=& \contraction{\tilde {J}^{(+)}(z) \tilde{J}^{(-)}(z')} z^{-\omega }V^{m+\omega}(z){z'}^\omega V^{n-\omega}(z') 
  \nonumber \\
  && +(m+\omega)(n-\omega)\contraction{\tilde{\psi}^{(+)}(z) \tilde{\psi}^{(-)}(z')}\psi_-(z)\psi_-(z')z^{-\omega }V^{m+\omega}(z){z'}^\omega V^{n-\omega}(z')
  \nonumber  \\
  &=&\Big [ \frac{1}{(z-z')^2}+\frac{\omega}{z'(z-z')}\Big ]z^{-\omega}V^{m+\omega}(z){z'}^{\omega}V^{n-\omega}(z')
  + \textrm{(fermion term = 0)}~, 
  \nonumber
\end{eqnarray} 
where $\contraction{\cdots}$ stands for the contraction. The residue at $z=z'$ is $(m+\omega)J_-V^{m+n}$.  Since $J_-V^{m+n}$ behaves like 
$\sim z^{m+n-1}$ near the origin, the final integral over $z'$ around the origin gives $(m+\omega)\delta _{m+n,0}$. This proves the formula of SGA 1. \\

\noindent
\textbf{SGA 2.} $\big\{\, B_r^{(\pm)}\,,~B_s^{(\mp)}~\big\} = \delta _{r+s,0}$ \\
\indent
Define
\begin{equation}
 B_r^{(\pm)} = \frac{1}{2\pi i} \oint dz B_r^{(\pm)}(z)~,
 \label{eq:6_intgral_from_Brpm}
\end{equation} 
where $B_r^{(\pm)}(z) =\Big[\psi^{(\pm)} \big(1+\half\psi_- \partial \psi_-J_-^{-2} \big) J_-^{1/2} - \psi_-J^{(\pm)} J_-^{-1/2} \Big] V^{r\pm\omega}$~. \\
\indent
The operator product expansion is given as
\begin{eqnarray}
 B_r^{(+)}(z)B_s^{(-)}(z') &=& \Big(1+\half\psi_-(z) \partial \psi_-(z)J_-^{-2}(z) \Big) J_-^{1/2}(z)V^{r + \omega}(z)z^{-\omega}
 \nonumber \\
 && \times \contraction{\tilde{\psi}^{(+)}(z)\tilde{\psi}^{(-)}(z')}{z'}^\omega V^{s-\omega}(z') 
\Big(1+\half\psi_-(z') \partial \psi_-(z')J_-^{-2}(z') \Big) J_-^{1/2}(z')
 \nonumber \\
 && +\psi_-(z)  J_-^{-1/2}(z)V^{r+\omega}(z) z^{-\omega}\contraction{\tilde{J}^{(+)}(z)\tilde{J}^{(-)}(z')}{z'}^{\omega}\psi_-(z')J_-^{-1/2}(z')
 V^{s-\omega}(z')
 \nonumber \\
 &=& \Big(1+\half\psi_-(z) \partial \psi_-(z)J_-^{-2}(z) \Big) J_-^{1/2}(z)V^{r + \omega}(z)z^{-\omega}
 \nonumber \\
 && \times \frac{1}{z-z'}{z'}^\omega V^{s-\omega}(z') \Big(1+\half\psi_-(z') \partial \psi_-(z')J_-^{-2}(z') \Big) 
 J_-^{1/2}(z')
 \nonumber \\
 && + \psi_-(z)  J_-^{-1/2}(z)V^{r+\omega}(z) z^{-\omega}\Big[\frac{1}{(z-z')^2}+\frac{\omega}{z'(z-z')}\Big]{z'}^{\omega}\psi_-(z')J_-^{-1/2}(z')
 V^{s-\omega}(z')
 \nonumber 
\end{eqnarray}
The radial ordering path integral over $z$ around $z'$
\begin{equation}
  \big\{\, B_r^{(\pm)}\,,~B_s^{(\mp)}(z')~\big\} = \frac{1}{2\pi i} \oint dz B_r^{(\pm)}(z) B_s^{(\mp)}(z')
\end{equation}
yields the residue
\begin{equation}
  \textrm{Residue} = V^{r+s}J_-\big(1+\psi_-\partial \psi_-J_-^{-2}\big) + \partial \psi_- J_-^{-1}V^{r+s}\psi_- = V^{r+s} J_-~.
  \nonumber
\end{equation}
The final integral over $z'$ around the origin gives $\delta _{r+s,0}$. \\
\indent
The following algebras, SGA 3-SGA 7, are proved by the same method as above, so we neglect the details. \\

\noindent
\textbf{SGA 3.} $\big[\, A_m^i\,,~B_r^j~\big] = 0$ \\

\noindent
\textbf{SGA 4.} $\big[\, A_m^{(\pm)}\,,~A_n^+~\big] = (m\pm\omega)A_{m+n}^{(\pm)}$ \\
\indent
The light-cone operator defined by
\begin{equation}
  A_n^{+} = \frac{1}{2\pi i} \oint dz A_n^+(z)~,
  \label{eq:6_intregral_from_An+}
\end{equation}
where $A_n^+(z) = \big[ \big(J_+ - n \psi_+ \psi_- \big) - \half n\big( \partial J_- J_-^{-1} -n\psi_- \partial \psi_- J_-^{-1} \big) \big]V^n$
is the same as Eq. (3. 13) in Ref.\cite{ref:Brower2}. \\

\noindent
\textbf{SGA 5.} $\big[\, B_r^{(\pm)}\,,~A_n^+~\big] = (\half n+ r \pm \omega)B_{r+n}^{(\pm)} $\\

\noindent
\textbf{SGA 6.} $\big[\, A_m^{(\pm)}\,,~B_r^+~\big] = (m \pm \omega ) B_{m+r}^{(\pm)} $\\
\indent
The light-cone operator defined by
\begin{equation}
  B_r^+ = \frac{1}{2\pi i} \oint dz B_r^+(z)~,
  \label{eq:6_interal_from_Br+}
\end{equation}
where $B_r^+(z) = \big[ \psi_+ \big(1+ \half\psi_- \partial \psi_- J_-^{-2}\big)J_-^{1/2} - \psi_- J_+ J_-^{-1/2} \big]V^r + \textrm{(- components)}$ \\
\indent
is the same as Eq. (3.15) in Ref.[7]. \\

\noindent
\textbf{SGA 7.} $\big\{\, B_r^{(\pm)}\,,~B_s^+~\big\} = A_{r+s}^{(\pm)}$ \\

\noindent
\textbf{SGA 8.} The commutativity of $G_r$ with $A_n^{(\pm)}$ \\
\indent
It is rather complicated to show the commutativity. So, let us give its detail in the following: \\
\noindent
Proof \\
Define $X(z) = \psi^{(\pm)}(z) V^{n\pm\omega}(z)$, which gives the $r$-independent expression  
\begin{equation}
  z^{-r-1/2}\big\{\, G_r\,, X(z)~\big\} = A_n^{(\pm)}(z) = \big[ J^{(\pm)} - (n\pm\omega)\psi^{(\pm)}\psi_-\big]V^{n\pm\omega}~,
  \label{eq:6_commutation_GrX}
\end{equation}
where
\begin{equation}
  A_n^{(\pm)} = \frac{1}{2\pi i} \oint dz A_n^{(\pm)}(z)~.
  \label{eq:6_integral_from_An}
\end{equation}
Then
\begin{equation}
  \big[\, G_r\,, A_n^{(\pm)}(z)~\big] = \big[\, G_r\,, z^{-r-1/2}\big\{\, G_r\,, X(z)~\big\}~\big]
  =z^{-r-1/2}\big[\, L_{2r}\,, X(z)~\big]~.
  \label{eq:6_commutation_GrX2}
\end{equation}
The operator product of $T_B(z)X(z')$ goes as follows (for only $(+)$ in $X(z)$):
\begin{eqnarray*}
  T_B(z)X(z') &=& \big[-:J_+(z)J_-(z): -\frac{\omega}{z}:\tilde {\psi}^{(+)}(z)\tilde {\psi}^{(-)}(z): + \half: \partial \tilde {\psi}^{(+)}(z)\tilde {\psi}^{(-)}(z):
  \\
  &&+ \half: \partial \tilde {\psi}^{(-)}(z)\tilde {\psi}^{(+)}(z): \big] 
    \tilde {\psi}^{(+)}(z')z'^{-\omega}V^{n+\omega}(z') 
  \\
  &=& -J_-(z)\contraction{J_+(z)V^{n+\omega}(z')}\tilde {\psi}^{(+)}(z')z'^{-\omega} 
 \\
  && +\big[-\frac{\omega}{z}\tilde {\psi}^{(+)}(z)\contraction{\tilde {\psi}^{(-)}(z)\tilde {\psi}^{(+)}(z')} 
  + \half \partial\tilde {\psi}^{(+)}(z) \contraction{\tilde {\psi}^{(-)}(z)\tilde {\psi}^{(+)}(z')} 
  \\
  &&-\half \tilde {\psi}^{(+)}(z)\contraction{\partial \tilde {\psi}^{(-)}(z)\tilde {\psi}^{(+)}(z')}\big] 
   z'^{-\omega}V^{n+\omega}(z')~.
\end{eqnarray*}
\begin{eqnarray*}
  \textrm{The first term} &=& -\psi^{+}J_-\frac{-1}{z-z'}(n+\omega)V^{n+\omega}=\frac{1}{z-z'}\psi^{(+)}\partial V^{n+\omega}~. \\
  \textrm{The second term} &=&  \big[-\frac{\omega}{z}\tilde {\psi}^{(+)}\frac{1}{z-z'} 
  + \half \partial\tilde {\psi}^{(+)} \frac{1}{z-z'}
  +\half \tilde {\psi}^{(+)}\frac{1}{(z-z')^2}\big]{z'}^{-\omega}V^{n+\omega}(z')~.
\end{eqnarray*}
The radial-ordering path integral is given by
\begin{equation}
  \big[\, L_{2r}\,, X(z'){(z')}^{-r-1/2}~\big] = \frac{1}{2\pi i}\oint dz z^{2r+1}T_B(z)X(z'){(z')}^{-r-1/2}~.
  \label{eq:6_comutation_rel_L2rXz}
\end{equation}
\begin{eqnarray*}
\textrm{The residue of the first term} 
 &=& \psi^{(+)}\partial V^{n+\omega} z^{r+1/2}~.
 \\
\textrm{The residue of the second term} 
 &=& \big[ -\frac{\omega}{z}\tilde {\psi}^{(+)} + \half \partial \tilde{\psi}^{(+)} \big]z^{-\omega}V^{n+\omega}z^{r+1/2} 
  \\
  && +\half \partial \big(z^{2r+1}\tilde{\psi}^{(+)})z^{-\omega}V^{n+\omega}z^{-r-1/2}
  \\
 &=&\partial \psi ^{(+)}V^{n+\omega}z^{r+1/2} + \psi ^{(+)}V^{n+\omega}\partial z^{r+1/2}~.
\end{eqnarray*}
Totally we have the total derivative given as Residue=$\partial ( \psi^{(+)}V^{n+\omega}z^{r+1/2})$. This shows that $G_r$ is commutative with 
$A_n^{(\pm)}$. \\

\noindent
\textbf{SGA 9.} The anticommutativity of $G_r$ with $B_s^{(\pm)}$ \\

\noindent
\textbf{Proof} \\
Define $Y(z)=\psi^{(\pm)}(z)\psi_-(z)J_-^{-1/2}(z)V^{s\pm\omega}(z)$, which gives the $r$-independent expression
\begin{equation}
 z^{-r-1/2}\big[\,G_r\,, Y(z)~\big]=B_s^{(\pm)}(z)=\Big[\psi^{(\pm)}\big(1+\half\psi_-\partial \psi_-J_-^{-2}\big)J_-^{1/2}
 -\psi_-J^{(\pm)}J_-^{-1/2}\Big]V^{s\pm \omega}~,
 \label{eq:6_del_Bspm}
\end{equation}
where
\begin{equation}
  B_s^{(\pm)} = \frac{1}{2\pi i} \oint dz B_s^{(\pm)}(z)~.
  \label{eq:6_integral_from_Bspm}
\end{equation}
Then
\begin{equation}
  \big\{\, G_r\,, B_s^{(\pm)}(z)~\big\} = \big\{\, G_r\,, z^{-r-1/2}\big[\, G_r\,, Y(z)~\big]~\big\}
  =z^{-r-1/2}\big[\, L_{2r}\,, Y(z)~\big]~.
  \label{eq:6_commutation_GrY2}
\end{equation}
The operator product of $T_B(z)Y(z')$ goes as follows (for only $(+)$ in $Y(z)$):
\begin{eqnarray*}
 T_B(z)Y(z') &=& \big[-:J_+(z)J_-(z): -\frac{\omega}{z}:\tilde {\psi}^{(+)}\tilde {\psi}^{(-)}: + \half : \partial \tilde {\psi}^{(+)}
 \tilde {\psi}^{(-)}: 
  \\ 
   &&+ \half: \partial \tilde {\psi}^{(-)}\tilde {\psi}^{(+)}: -\half: \partial \psi_{+}\psi_{-}: - \half: \partial\psi_{-}\psi_{+}:\big]
   \\
   &&\times \psi^{(+)}(z'){z'}^{-\omega}\psi_-(z')J_-^{-1/2}(z')V^{s+\omega}(z') \\
 &=& -J_-(z)\frac{1}{2(z-z')}J_-^{-3/2}V^{s+\omega}\psi^{(+)}\psi_-  +\frac{s+\omega}{z-z'}J_-^{1/2}V^{s+\omega}\psi^{(+)}\psi_- \\
  &&+\Big[-\frac{\omega}{z}\tilde{\psi}^{(+)}\frac{1}{z-z'}+\half\partial\tilde{\psi}^{(+)}\frac{1}{z-z'}+\half \tilde{\psi}^{(+)}
 \frac{1}{(z-z')^2}\Big]{z'}^{-\omega}\psi_- J_-^{-1/2}V^{s+\omega}(z') \\
 &&+\Big[ -\half \psi_-(z)\frac{1}{(z-z')^2} - \half \partial \psi_- \frac{1}{z-z'}\Big]\psi^{(+)}J_-^{-1/2}V^{s+\omega}~.
\end{eqnarray*}
The radial-ordering path integral is given by
\begin{equation}
  \big[\, L_{2r}\,, Y(z'){(z')}^{-r-1/2}~\big] = \frac{1}{2\pi i}\oint dz z^{2r+1}T_B(z)Y(z'){(z')}^{-r-1/2}~.
  \label{eq:6_comutation_rel_L2rYz}
\end{equation}

Finally we have the total derivative given as Residue $=\partial (\psi^{(+)}\psi_-J_-^{-1/2}V^{s+\omega}z^{r+1/2})$. This shows that $G_r$ is anticommutative with $B_s^{(\pm)}$. \\

\noindent
\textbf{SGA 10.}  The commutativity of $G_r$ with $A_n^+$, $B_s^+$ \\
\indent 
Since $A_n^+$, $B_s^+$ are all composed of the light-cone variables $\pm$, $G_r$ are already known to be commutative with $A_n^+$, $B_s^+$. The terms related to $\omega$ in $G_r$ are irrelevant in this proof. \\

%
%
%
%
\section{Different D-branes}  \label{sec:6}

We consider two different D-branes, where the charged open string is attached to the $Dp$-brane and $Dp'$-brane ($p>p'$,
 they are all even numbers) at $\sigma = 0$  and $\sigma = \pi$, respectively.
 The magnetic fields are introduced in such a way that they are diagonalizable into two-dimensional blocks
 in subspaces $\mu = 1, 2, \cdots, p'$ and $\mu =p'+1, \cdots, p$, respectively. Thus, for the first subspace $\mu = 1, 2, \cdots, p'$,
 the situation is the same as in the previous sections, namely, we have the same (NN) boundary conditions such as
 $X'^\mu = -\rho (\sigma )F^\mu_{\ \nu}\dot{X}^\nu$ at $\sigma =0$ and $\pi$, (Neumann), while, in the second subspace $\mu =p'+1, \cdots, p$,
 we have the (ND) mixed boundary conditions such as $X'^\mu = -q_0F^\mu_{\ \nu}\dot{X}^\nu$ at $\sigma = 0$, (Neumann),
 and $X^\mu = c_\pi^\mu($const.$)$ at $\sigma =\pi$ (Dirichlet).  \\
\indent
The normal mode expansions for $X^\mu(\tau, \sigma)$ corresponding to the ND boundary conditions are
\begin{equation}
  X^{(\pm)} = c_\pi^{(\pm)} \pm i\sum_r \frac{1}{r \pm \omega_0} e^{-i(r \pm \omega_0)\tau }\cos \big[ (r \pm \omega _0)\sigma \mp \pi \omega _0 \big] 
  \alpha _r^{(\pm )}~,
 \label{eq:6_normal_mode_Xpm}
\end{equation}
where $r$ runs over half-integers, and $\tan \pi\omega_0 = q_0 B$, $|\omega_0| < 1/2$ is a parameter of the $i$-th block. As in Sec.3, we choose the coordinate axes suitably to make $\omega_0$ positive. These mode expansions are just opposite from the conventional integer ones.\\
\indent
For the DD boundary conditions, we have
\begin{equation}
  X^a = c_0^a + q^a\sigma + \sum_{n\neq 0}\frac{1}{n} e^{-in\tau } \alpha _n^a \sin (n\sigma )~,
  \label{eq:6_DD_bc}
\end{equation}
where $n$ runs over integers, and $c_0^a$, $c_0^a + q^a \pi = c_\pi^a$ are constants, indicating the locations of $D$-branes. \\
\indent
The fermionic mode expansions corresponding to Eq.(\ref{eq:6_normal_mode_Xpm}) become also opposite from the conventional ones \cite{ref:Arfaei}. This comes from the supertransformation, $\delta X^\mu=\bar {\epsilon } \psi^{\mu}$.  The NS sector corresponds to the half-integer expansion of $\bar{\epsilon}(z)$, which leads to the integer expansion of $\psi(z)$ as
\begin{equation}
  \psi_{\rm{NS}}^{(\pm)}(z) = \sum_n b_n^{(\pm)} z^{-n\mp \omega_0 -1/2} = z^{\mp \omega_0} \tilde {\psi}_{\rm{NS}}^{(\pm)}(z)~,
  \label{eq:6_integar_mode_psi}
\end{equation}
where $n$ runs over integers. The R sector corresponds to the integer expansion of $\bar{\epsilon }(z)$, which leads to the half-integer expansion of $\psi(z)$ as
\begin{eqnarray}
  \psi_{\rm{R}}^{(\pm)}(z) = \sum_r d_r^{(\pm)} z^{-r\mp \omega_0 - 1/2} = z^{\mp \omega_0}\tilde{\psi}_{\rm{R}}^{(\pm)}(z)
  \label{eq:6_expansion_psiR}
\end{eqnarray}
where $r$ runs over half-integers. \\
\indent
We now consider the system obeying the ND boundary conditions. For the $i$-th block, the operator product expansions are given by
\begin{equation}
  \tilde {J}^{(\pm)}(z) \tilde {J}^{(\mp)}(z') = \Big[\frac{z+z'}{2(z-z')^2} \pm \frac{\omega_0}{(z-z')}\Big]\sqrt{zz'}~,
  \label{eq:6_op_product_ex_J_J}
\end{equation}
with $\tilde {J}^{(\pm)}(z) =\sum_r \alpha _r^{(\pm)} z^{-r}$, and
\begin{eqnarray}
  \tilde {\psi}^{(+)}_{\rm{NS}}(z) \tilde {\psi}^{(-)}_{\rm{NS}}(z') &=& \frac{z}{z-z'}\frac{1}{\sqrt{zz'}} \qquad 
  \tilde {\psi}^{(-)}_{\rm{NS}}(z) \tilde {\psi}^{(+)}_{\rm{NS}}(z') = \frac{z'}{z-z'}\frac{1}{\sqrt{zz'}}~, 
  \label{eq:6_op_product_ex_psiNS_psiNS} \\
  \tilde {\psi}^{(+)}_{\rm{R}}(z) \tilde {\psi}^{(-)}_{\rm{R}}(z') &=& \frac{1}{z-z'}~, 
  \label{eq:6_op_product_ex_psiR_psiR} 
\end{eqnarray} 
Then we have, for the NS sector,
\begin{eqnarray*}
 T_F(z) T_F(z') &=& \big[\tilde {J}^{(+)}(z) \tilde {\psi}^{(-)}(z) + \tilde {J}^{(-)}(z) \tilde {\psi}^{(+)}(z) \big]
                  \big[\tilde {J}^{(+)}(z') \tilde {\psi}^{(-)}(z') + \tilde {J}^{(-)}(z') \tilde {\psi}^{(+)}(z') \big]
                \\
                &=& \big[\frac{1}{(z-z')^3} + \frac{\omega_0}{(z-z')^2}\big] z' 
                +\big[\frac{1}{(z-z')^3} - \frac{\omega_0}{(z-z')^2}\big] z 
                +\frac{2}{z-z'} T_B(z') 
                \\
                &=&\frac{(z+z')^2}{2(z-z')^3} - \frac{\omega_0}{(z-z')} +\frac{2}{z-z'}T_B(z')~. 
\end{eqnarray*}
This is the formula for 2-dimensions of the $i$-th block. Since the number of the block is $(p-p')/2$, we have totally
\begin{equation}
  T_F(z) T_F(z') = \frac{p-p'}{2(z-z')^3}\frac{(z+z')^2}{2} + \frac{\sum_{i=ND} \omega_{0i}}{z-z'} +\frac{2}{z-z'}T_B(z')
  \label{eq:6_total_TBTB}
\end{equation}
This is equivalent to the algebra for the $(p-p')$ dimensions,   
\begin{equation}
  \big\{\, G_r\,,~G_s~\big\} = 2L_{r+s} + B^{p-p'}(r)\delta _{r+s,0}~,
  \label{eq:6_commutation_rel_BrBr2}
\end{equation}
where
\begin{equation}
  B^{p-p'}(r) = \frac{p-p'}{2}\Big(r^2 + \frac{1}{4} \Big) - \sum_{i=ND} \omega_{0i}~.
  \label{eq:6_B(p-p')}
\end{equation}

 In the following we summarize cyclotron frequencies involved in the NS and R sectors in various dimensions:\\

\begin{tabular}{lcccc} \hline
                          &   Bound. Cond. & NS       & R       & cycl. freq.                  \\ \hline 
I. $\mu=1,\cdots,p':$      & N-N            & $(n,r)$ & $(n,n)$ & $\omega=\omega_0-\omega_\pi$ \\
II. $\mu=p'+1,\cdots,p:$  & N-D            & $(r,n)$  & $(r,r)$ & $\omega_0$                   \\ 
III. $\mu=p+1,\cdots,d-2:$ & D-D           & $(n,r)$  & $(n,n)$ & $0$                          \\
IV. $\mu=0,d-1:$           & N-N           & $(n,r)$  & $(n,n)$ & $0$                          \\ \hline
\end{tabular}
\\
\\
\noindent
Here, N and D stand for the Neumann and the Dirichlet boundary conditions. The ($n$, $r$) corresponds to the normal mode expansions of ($X$, $\psi$), where $n$ takes integers and $r$ takes half-integers. The cyclotron frequencies are generally involved in ( $X$, $\psi$ ) in I and II, but not in III and IV.  However, for the R sector, no cyclotron frequencies are involved in anomaly factors $A$, $B$.\\
\indent
As for IV, the construction of SGA requires that the light-cone dimensions are subject to the NN boundary conditions. \\
\indent
Each anomaly in I, II, III, IV is given by \\

\noindent
I. 
\begin{eqnarray*}
  A^{\rm{I}}_{\rm{R}}(m)&=&\frac{1}{8}p'm^3~, \qquad \qquad B^{\rm{I}}_{\rm{R}}(m)=\frac{1}{2}p'm^2~,  \\
  A^{\rm{I}}_{\rm{NS}}(m)&=&\frac{1}{8}p'm(m^2-1)+m\omega^{\rm{I}}~, \qquad B^{\rm{I}}_{\rm{NS}}(r)=\frac{1}{2}p'(r^2-\frac{1}{4})+\omega^{\rm{I}}~,  \\
  \omega^{\rm{I}} &=& \sum_{i\in \rm{I}}\omega_{i}~.
\end{eqnarray*}
II.
\begin{eqnarray*}
  A^{\rm{II}}_{\rm{R}}(m)&=&\frac{1}{8}(p-p')m^3~, \qquad \qquad B^{\rm{II}}_{\rm{R}}(m)=\frac{1}{2}(p-p')m^2~,  \\
  A^{\rm{II}}_{\rm{NS}}(m)&=&\frac{1}{8}(p-p')m(m^2+1)-m\omega _0^{\rm{II}}~, \\
  B^{\rm{II}}_{\rm{NS}}(r)&=&\frac{1}{2}(p-p')(r^2+\frac{1}{4}) - \omega _0^{\rm{II}}~,  \\
  \omega_0^{\rm{II}} &=& \sum_{i\in \rm{II}}\omega_{0i}~.
\end{eqnarray*}
III.
\begin{eqnarray*}
  A^{\rm{III}}_{\rm{R}}(m)&=&\frac{1}{8}(d-2-p)m^3~, \qquad \qquad B^{\rm{III}}_{\rm{R}}(m)=\frac{1}{2}(d-2-p)m^2~,  \\
  A^{\rm{III}}_{\rm{NS}}(m)&=&\frac{1}{8}(d-2-p)m(m^2-1)~, \qquad B^{\rm{III}}_{\rm{NS}}(r)=\frac{1}{2}(d-2-p)(r^2-\frac{1}{4})~.
\end{eqnarray*}
IV.
\begin{eqnarray*}
  A^{\rm{IV}}_{\rm{R}}(m)&=&\frac{1}{4}m^3~, \qquad \qquad B^{\rm{IV}}_{\rm{R}}(m)=m^2~,  \\
  A^{\rm{IV}}_{\rm{NS}}(m)&=&\frac{1}{4}m(m^2-1)~, \qquad B^{\rm{IV}}_{\rm{NS}}(r)=r^2-\frac{1}{4}~.
\end{eqnarray*}  
By adding all together we have
\begin{eqnarray}
  A_{\rm{R}}(m)&=&\frac{1}{8}dm^3~, \qquad B_{\rm{R}}(m)=\frac{1}{2}dm^2~,  
  \nonumber \\
  A_{\rm{NS}}(m)&=&\frac{1}{8}dm(m^2-1)+\frac{1}{4}(p-p')m+m \omega^{\rm{I}}-m\omega_0^{\rm{II}}~, 
  \label{eq:6_value_A_B} \\
  B_{\rm{NS}}(r)&=&\frac{1}{2}d(r^2-\frac{1}{4})+\frac{1}{4}(p-p')+\omega^{\rm{I}}-\omega_0^{\rm{II}}~. 
  \nonumber
\end{eqnarray}
In the case of no $D$-branes, we should set $p=p'=d-2$, and $\omega_0^{\rm{II}}=0$.

The Regge intercept is defined as follows: The 0-th Virasoro operator is written as
\begin{equation}
  L_0 = R + \frac{1}{2}\big( p^2 + q^2 \big) = a~,
  \label{eq:6_def_Regge_intercept}
\end{equation}
where $a$ is a constraint constant, $R$ is the oscillator operator, and ($p$, $q$) is the displacement operator of ($\tau $, $\sigma $),  in the mode expansions of coordinates, respectively. If we set $p^2=-s$, we have $R=\frac{1}{2}s+a-\frac{1}{2}q^2$. This means that the Regge intercept is given by
\begin{equation}
  \alpha (0) = a - \frac{1}{2}q^2
  \label{eq:6_value_alpha0}
\end{equation}
where $q^2=\sum_{i=p+1}^{d-2}(q_i)^2$, $q_i$ is a distance between two $D$-branes.  \\

The isomorphism (\ref{eq:4_similar_A_B}) and hence (\ref{eq:4_condition_AT_BT}) now turns to
\begin{equation}
   B_{\rm{NS}}^T(r) = \frac{d-2}{2}(r^2-\frac{1}{4}) + 2a +\frac{1}{4}(p-p') + \omega^{\rm{I}} - \omega_0^{\rm{II}} = 4r^2~,
   \label{eq:6_def_Br2}
\end{equation}
so that 
\begin{equation}
  a = \half \big[ 1 - \frac{1}{4}(p-p') - (\omega^{\rm{I}} - \omega^{\rm{II}}_0) \big]~, \qquad d=10~.
  \label{eq:6_value_a}
\end{equation}
Hence we get, for the NS sector,
\begin{equation}
   \alpha_{\rm{NS}}(0) = \half \big[ 1 - \frac{1}{4}(p-p') - \omega^{\rm{I}} + \omega^{\rm{II}}_0 - q^2 \big]~,
  \label{eq:6_condition_alpha_NS}
\end{equation}
In the same way we have $a=0$ for the R sector, hence
\begin{equation}
  \alpha_{\rm{R}}(0) =  - \half q^2 ~.
  \label{eq:6_def_alpha0_R}
\end{equation}

\section{Concluding remarks}

We have constructed the SGA for charged superstrings placed in constant background magnetic fields. Contrary to the neutral string this algebra
 is characteristic of including the cyclotron frequency $\omega$. Any physical state satisfying the super Virasoro condition can be constructed
 from elements of this SGA, if the space-time dimension is $d=10$ and the Regge intercept $\alpha (0)$ is given by
  $\alpha _{\rm{NS}}(0)=(1-\omega)/2$ for the NS sector,
 and $\alpha_{\rm{R}}(0)=0$ for the R sector, when there is no $D$-brane. More generally, when there are $Dp$- and $Dp'$-branes,
 the Regge intercept is given by Eq.(\ref{eq:6_condition_alpha_NS}) for the NS sector and Eq.(\ref{eq:6_def_alpha0_R}) for the R sector.
 Superconformal anomalies are also given by Eqs.(\ref{eq:3_del_A_B}), (\ref{eq:3_del_Br2}) and (\ref{eq:6_value_A_B}). \\
\indent
Generally, the mass spectrum under the magnetic field is given by the formula, for the NN case,
\begin{equation}
  \alpha 'M^2 = R -a~,
  \label{eq:7_mass_spectrum_2}
\end{equation}
where $a = (1-\omega )/2$ for the NS sector and $a =0$ for the R sector. The number operator $R$ can be written as
\begin{equation}
  R = R_0 + \sum_i N_i \omega_i~.
  \label{eq:7_R_form2}
\end{equation}
Here $R_0$ is the number operator taking eigenvalues $0, 1/2, 1, 3/2, 2, \cdots $for the NS sector, and $0, 1, 2, \cdots $ for the R sector, whereas $N_i$ takes values $0$ and  $\pm$integral values. The negative integral values of $N_i$ are restricted by the lower bounds
 $\alpha' M^2 > -1/2$ for the NS sector, and $\alpha' M^2>0$ for the R sector. \\
\indent
To sum up, the mass-spectrum formula can be expressed as
\begin{eqnarray}
  \alpha' M^2_{\rm{R}} &=& R_{\rm{R}} +  \sum_i N_i \omega_i \qquad \qquad R_{\rm{R}}=0, 1, 2, \cdots, \quad \textrm{ for the R sector}~,
  \label{eq:7_M2R_form3} \\
  \alpha' M^2_{\rm{NS}} &=& R_{\rm{NS}} + \half \omega  + \sum_i N_i \omega_i \quad R_{\rm{NS}}=-\half, 0, \half, 1, \frac{3}{2}, 2, \cdots \quad \textrm{for the NS sector}~, \nonumber \\
  \label{eq:7_M2NS_form3}
\end{eqnarray}
where $\omega = \sum_i \omega_i$. The GSO projection may be useful, if half-integral values, -1/2, 1/2, 3/2, care projected out of $R_{NS}$. However, the constant factor $\omega /2$ still remains. This means we cannot obtain the space-time supersymmetric model.

\section*{Acknowledgement}
It is a pleasure to thank T. Okamura for continuous discussions. 


\end{document}